\documentclass{article}
\usepackage{spconf,amsmath,amssymb,graphicx,booktabs}
\usepackage{soul}
\usepackage{graphicx,color}
\usepackage{textcomp}
\usepackage{xcolor}
\usepackage{siunitx}

\DeclareMathOperator*{\argmin}{arg\,min}

\title{Complex Domain Inverse Problem Formulation for 4D Flow MRI Super-Resolution}
\name{Aurélien de Turenne\textsuperscript{1}, Denis Kouamé\textsuperscript{1}}
\address{\textsuperscript{1} Univ Toulouse, Toulouse INP, CNRS, IRIT, Toulouse, France}

\begin{document}
\raggedbottom
\maketitle

\begin{abstract}
Four-dimensional flow magnetic resonance imaging measures blood velocity fields in three dimensions throughout the cardiac cycle and is increasingly used to assess hemodynamics. However, these measurements are limited by low signal-to-noise ratios and limited spatial resolution, which can compromise their accuracy and reliability. To address these limitations, we investigate the feasibility of spatial resolution enhancement through a complex-domain inverse problem formulation inspired by MRI acquisition principles. Three inversion methods are evaluated: a closed-form approach (FSR), a Plug-and-Play alternating direction method of multipliers (PnP-ADMM), and a pretrained network that takes the forward operator as input (RAM). On simulated data, all three methods outperform 4DFlowNet at $\times 2$ and $\times 4$ super-resolution factors. On aortic phantom data at $\times 2$ super-resolution, PnP-ADMM and RAM achieve lower root-mean-square error than 4DFlowNet in almost all cases. These results highlight the potential of the proposed formulation and support further development and validation.
\end{abstract}


\section{Introduction}
Four-dimensional phase-contrast MRI, also known as 4D Flow MRI, measures the three components of blood velocity over a 3D volume and the cardiac cycle \cite{markl_4d_2012}. It enables qualitative visualization through streamlines as well as quantitative analysis through the estimation of biomarkers such as wall shear stress or relative pressure. Such biomarkers support several clinical applications, for instance the prediction of aortic dissection evolution \cite{takahashi_utility_2022, adriaans_predictive_2019}. However, acquisition-time constraints in clinical practice limit spatial resolution and signal-to-noise ratio \cite{bissell_4d_2023}. To address these limitations, spatial super-resolution methods have been proposed to improve velocity-field representation without increasing acquisition duration.

Supervised deep-learning methods such as 4DFlowNet \cite{ferdian_4dflownet_2020} learn spatial super-resolution from paired low and high resolution velocity fields generated by computational fluid dynamics (CFD), as acquiring such pairs \textit{in vitro} or \textit{in vivo} is difficult. Reliance on synthetic training data raises questions about generalization to unseen anatomies, particularly in pathological cases such as in aortic dissections.

Alternatively, Levilly et al. \cite{levilly_segmentation-free_2024} formulate super-resolution as an inverse problem with Navier-Stokes-based regularization. That formulation is posed directly in the velocity domain and relies on an empirical forward model rather than one derived from the physics of MRI acquisition.

In this paper, we propose a complex-signal domain inverse-problem formulation using a forward model derived directly from MRI acquisition principles, where velocity is extracted only after reconstruction. To avoid reliance on paired training data, we restrict ourselves to approaches requiring no task-specific training, across three paradigms: an analytical closed-form called Fast Super-Resolution (FSR)  \cite{tuador_novel_2021}, a Plug-and-Play alternating direction method of multipliers (PnP-ADMM) \cite{venkatakrishnan_plug-and-play_2013, zhang_plug-and-play_2021}, and a fully pretrained reconstruction network named Reconstruct Anything Model (RAM) \cite{terris_reconstruct_2025}.

Our contribution is to formulate 4D Flow MRI super-resolution as a complex-domain inverse problem and assess its effectiveness using different inversion methods, benchmarking their performance against the supervised baseline 4DFlowNet on both simulated and phantom data. The paper is organized as follows: Section \ref{sec:method} presents the acquisition forward model, the inverse problem formulation and the three approaches used to solve it. Section \ref{sec:results} details the experimental setup and discusses the results. Section \ref{sec:conclusion} concludes the paper.

\section{Methods}\label{sec:method}
We first formulate the physics-based forward model in the complex domain, then describe the generic regularized inverse problem and the three selected methods to solve it.

Throughout the paper, we consider an arbitrary but fixed cardiac time frame $t$ and omit the time index for readability. Reconstruction is performed independently for each frame and velocity component $j \in \{u,v,w\}$.

\subsection{Acquisition process and forward model}
In 4D Flow MRI, blood velocity is measured by applying velocity-encoding bipolar gradients along three orthogonal directions. At a fixed time frame, the acquisition consists of a reference signal $s_{\mathrm{ref}}=A e^{i\Phi_{\mathrm{ref}}}$ and three encoded signals $s_{\mathrm{venc},j}=A e^{i(\Phi_{\mathrm{ref}}+\Phi_j)}$, where $j\in\{u,v,w\}$ denotes the velocity component. The scanner samples these signals in $k$-space and reconstructs them by inverse Fourier transform. The phase $\Phi_j$ is related to velocity $\nu_j$ by
\begin{equation}
\Phi_j=\pi\frac{\nu_j}{\mathrm{VENC}_j},
\label{eq:venc-relation}
\end{equation}
with VENC the velocity corresponding to a phase shift of $\pi$.

From the magnitude image $A$ and velocity $\nu_j$, we recover the phase $\Phi_j$ via Eq. \eqref{eq:venc-relation} and construct a complex signal for each encoding direction:
\begin{equation}
s_j = A\, e^{i\Phi_j}, \qquad j \in \{u,v,w\}.
\label{eq:complex-signal}
\end{equation}
The acquisition spatial resolution is determined by the size of the sampled $k$-space. A downsampling by $\delta$ on each axis is modeled as retaining a centered block containing $1/\delta^3$ of the full $k$-space. In the spatial domain, this operation is equivalent to convolution with a three-dimensional sinc kernel followed by decimation by a factor of $\delta$ along each axis.

Let $y_j$ be the lexicographic vectorization of the low-resolution (LR) image $s_j$, and $x_j$ the unknown high-resolution (HR) signal. We then write this frequency truncation in the spatial domain as
\begin{equation}
y_j = SH x_j + \varepsilon_j, \qquad j \in \{u,v,w\},
\label{eq:forward}
\end{equation}
where $y_j \in \mathbb{C}^{N_l}$ ($N_l = m_l n_l p_l$) and $x_j \in \mathbb{C}^{N_h}$ ($N_h = m_h n_h p_h$). $H \in \mathbb{R}^{N_h \times N_h}$ is a 3-level block circulant matrix of the sinc circular convolution. Its Fourier decomposition is $H=F^{H}\Lambda F$, where $F$ is the unitary 3D DFT matrix and $\Lambda$ is the diagonal matrix of eigenvalues of $H$. $S \in \mathbb{R}^{N_l \times N_h}$ decimates with per-axis rates $d_X, d_Y, d_Z$ (we consider  in all experiments, without loss of generality the isotropic case where, $d_X=d_Y=d_Z=\delta$), and $\varepsilon_j$ is additive complex white Gaussian noise.

Working directly on $s_j$, rather than on the phase, keeps this degradation linear and the noise additive Gaussian. Once an HR estimate $\hat{x}_j$ is recovered, its phase is extracted and converted to velocity using Eq. \eqref{eq:venc-relation}.

\subsection{Inverse problem}
Recovering $x_j$ from $y_j$ in Eq. \eqref{eq:forward} is an ill-posed problem. We therefore estimate $x_j$ as the solution of a regularized variational problem,
\begin{equation}
\hat{x}_j \in \argmin_{x_j}\;
\frac{1}{2}\,\lVert y_j - SH x_j \rVert_2^2 + \tau\, R(x_j),
\label{eq:generic-inverse}
\end{equation}
for $j \in \{u,v,w\}$, where the quadratic term enforces consistency with the forward model, $R(\cdot)$ encodes prior knowledge on the HR signal, and $\tau > 0$ is a scalar controlling the relative weight of the two terms.

For FSR and PnP-ADMM, regularization parameters are selected per component by grid search against the full simulation ground truth, then reused on the phantom. RAM requires no such tuning. Whatever method is used to solve Eq. \eqref{eq:generic-inverse}, its outcome is an estimate $\hat{x}_j$ of the HR \emph{complex} signal. The velocity component along direction $j$ is then recovered by extracting the phase and applying Eq. \eqref{eq:venc-relation}. In the descriptions of the methods below, we also omit $j$, with $x$ and $y$ referring to a single component at a fixed frame. In addition, for the sake of readability, we present below the main steps of these algorithms.

\subsubsection{Fast Super-Resolution (FSR)}\label{sec:method:fsr}

With the Tikhonov prior $R(x)=\lVert x-\bar{x}\rVert_2^2$, where $\bar{x}$ is a bicubic interpolation estimate, problem \eqref{eq:generic-inverse} has the closed-form solution
\begin{equation}
\hat{x} = \big( H^{H} S^{H} S H + 2\tau I_{N_h} \big)^{-1}
\big( H^{H} S^{H} y + 2\tau \bar{x} \big),
\label{eq:fsr-normal}
\end{equation}
where $\cdot^{H}$ denotes the Hermitian transpose. Although, this solution requires inverting a large $N_h \times N_h$ matrix, which is impractical. FSR, extended to 3D in \cite{tuador_novel_2021}, avoids large matrix inversion and iterative optimization through the expression:
\begin{equation}
\hat{x} = \frac{1}{2\tau}\, g
- \frac{1}{2\tau}\, F^{H} \underline{\Lambda}^{H}
\big( 2\tau \delta^3 I_{N_l} + \underline{\Lambda}\, \underline{\Lambda}^{H} \big)^{-1}
\underline{\Lambda}\, F\, g ,
\label{eq:fsr-closed}
\end{equation}
where $g=H^{H}S^{H}y+2\tau\bar{x}$ and
\begin{equation}
\underline{\Lambda} =
\Big(
\big( \mathbf{1}_{d_Z}^{T} \otimes I_{p_l} \big)
\otimes
\big( \mathbf{1}_{d_Y}^{T} \otimes I_{n_l} \big)
\otimes
\big( \mathbf{1}_{d_X}^{T} \otimes I_{m_l} \big)
\Big)\,
\Lambda,
\label{eq:lambda-underline}
\end{equation}

with $\underline{\Lambda} \;\in\; \mathbb{C}^{N_l \times N_h}$, $\mathbf{1}_{\alpha}^{T} \in \mathbb{R}^{1 \times \alpha}$ is a row vector of ones, $I_\beta \in \mathbb{R}^{\beta \times \beta}$ the identity matrix, and $\otimes$ the Kronecker product. FSR serves both as a standalone reconstruction method and as a method to solve the data-consistency step within PnP-ADMM.

\subsubsection{Plug-and-Play ADMM with a pretrained denoiser (PnP-ADMM)}\label{sec:method:pnp}
To use a stronger prior while keeping the efficiency of FSR, we solve \eqref{eq:generic-inverse} with PnP-ADMM, in which the ADMM step involving $R$ is replaced by denoising with a pretrained network, so that $R$ never needs to be written explicitly. Introducing an auxiliary variable $z=x$, the scaled-form augmented Lagrangian comes to
\begin{equation}
\mathcal{L}(x,z,\eta) =
\tfrac{1}{2}\,\lVert y - SH x \rVert_2^2 + \tau\, R(z)
+ \tfrac{\mu}{2}\,\lVert z - x + \eta \rVert_2^2 ,
\label{eq:pnp-lagrangian}
\end{equation}
where $\eta$ is the scaled dual variable and $\mu > 0$ the penalty parameter. Minimizing $\mathcal{L}$ by ADMM gives the alternating updates:
\begin{align}
x^{k+1} &= \argmin_{x}\; \tfrac{1}{2}\lVert y-SHx\rVert_2^2 + \tfrac{\mu}{2}\lVert z^{k}-x+\eta^{k}\rVert_2^2,
\label{eq:pnp-x}\\
z^{k+1} &= \argmin_{z}\; \tau\, R(z)
+ \tfrac{\mu}{2}\,\lVert z - x^{k+1} + \eta^{k} \rVert_2^2 ,
\label{eq:pnp-z}\\
\eta^{k+1} &= \eta^{k} + z^{k+1} - x^{k+1} .
\label{eq:pnp-d}
\end{align}

The $x$-update of Eq. \eqref{eq:pnp-x} is a Tikhonov problem and is solved in closed form by FSR using Eq. \eqref{eq:fsr-closed}, applied to $y$ with $z^k+\eta^k$ as reference estimate and $\mu/2$ as regularization weight. The core idea of Plug-and-Play is that the $z$-update in Eq. \eqref{eq:pnp-z} is a denoising step, and can therefore be replaced by a denoiser $D_{\sigma_k}$ applied to $x^{k+1}-\eta^k$, without ever specifying $R$ explicitly, where $\sigma_k$ is the denoising level. Here $D_{\sigma_k}$ is the pretrained DRUNet denoiser \cite{zhang_plug-and-play_2021}, applied to the real and imaginary parts separately, component by component and frame by frame, each volume processed slice by slice along the sagittal direction.

\subsubsection{Reconstruct Anything Model (RAM)}\label{sec:method:ram}

RAM is a fully pretrained reconstruction network combining a 2D DRUNet with a Krylov subspace module. Unlike a supervised reconstructor such as 4DFlowNet, it takes the explicit forward operator as an input alongside the measurements (being real or complex), and is therefore applicable to a new degradation without retraining. We use the released pretrained weights without task-specific training. 

Nevertheless, RAM only handles 2D images while our problem involves 3D images degraded by a 3D operator. It cannot be applied directly. Therefore, we exploit the separability of the sinc convolution into in-plane and through-plane operators,
\begin{equation}
H = H_Z H_{XY}, \qquad \text{with } H_{XY} = H_X H_Y,
\label{eq:separable-h}
\end{equation}
where $H_X$, $H_Y$, and $H_Z$ are 1D convolutions along the respective axes. Decimation similarly separates as $S=S_ZS_{XY}$. We reconstruct each complex volume in two passes. First, RAM super-resolves axial slices along both in-plane axes, producing high in-plane but low through-plane resolution. Then, slices containing the through-plane axis are super-resolved only along that axis, using an operator restricted to that direction. Together, these passes map the LR signal $y$ to the HR estimate $\hat{x}$.

\section{Results}\label{sec:results}
In this section, we describe the datasets and evaluation metric, then report quantitative and qualitative results before discussing the limitations.

\subsection{Data description and metric}
We evaluate our methods on two datasets: a simulated CFD dataset and a real MRI acquisition of a flow phantom.

\textbf{Simulation.} The 4DFlowNet paper by Ferdian \textit{et al.} \cite{ferdian_4dflownet_2020}, to which we refer for further details on the simulation dataset, introduced a CFD-based synthetic 4D Flow MRI dataset derived from three real aortic geometries. Two geometries were used for training, while the third, \texttt{aorta03}, was reserved for evaluation. To ensure a direct comparison with 4DFlowNet, we use this same \texttt{aorta03} case. The CFD models a full cardiac cycle discretized into $N_t = 71$ time frames and provides HR velocity fields with isotropic voxel size $0.594 \times 0.594 \times 0.594\text{~mm}^3$, from which LR complex data are generated at decimation factors $\delta=2$ and $\delta=4$.

\textbf{Phantom.} We use the rigid aortic model from the \textit{in vitro} dataset by Zimmermann \textit{et al.} \cite{zimmermann_impact_2021} to which we refer for more details on the physical phantom and acquisition setup. The LR volume is a real 3D MRI acquisition discretized into $N_t = 50$ time frames per cardiac cycle, with isotropic voxel size $2.5 \times 2.5 \times 2.5\text{~mm}^3$. The reference ground truth consists of seven HR co-registered 2D phase-contrast (PC) measurements along anatomical planes, with in-plane resolution $1.1 \times 1.1\text{~mm}^2$. To obtain a pixel-to-pixel correspondence, super-resolved volumes are interpolated onto the 2D PC grid prior to comparison. We retain five of these planes: AAO\_RECOIL, BCT\_RECOIL, BRANCH\_RECOIL, INLET, and LSA, excluding the remaining two due to phase aliasing. Each slice intersects two vascular lumen cross-sections, denoted C1 and C2 (see Figure \ref{fig:phantom}), except LSA which contains a single one.

\textbf{Metric.} We compare FSR, PnP-ADMM, and RAM with 4DFlowNet. Evaluation is measured only inside vascular masks (i.e. discarding background), using the root-mean-square error (RMSE) between the reconstructed and reference velocity for each component, computed over all spatial samples and all cardiac time frames. Simulation errors are computed per velocity component over the 3D flow region. Phantom errors are computed, for each plane, between the through-plane velocity of the reconstruction and the 2D PC reference.

\subsection{Simulation and phantom evaluation}

Table \ref{tab:simu-rmse} reports the masked RMSE of each velocity component for $\delta=2$ and $\delta=4$. All three inverse-problem-based methods outperform 4DFlowNet on every component and at both factors, and the gap increases with the decimation factor. RAM shows the best results, closely followed by the PnP approach.

\begin{table}[h]
\caption{Masked RMSE (cm/s) per velocity component on the \texttt{aorta03} simulation dataset at $\times2$ and $\times4$ super-resolution, over 71 cardiac time frames. Bold marks the lowest error per row.}
\label{tab:simu-rmse}
\centering\small
\begin{tabular*}{\linewidth}{@{\extracolsep{\fill}}ccS[table-format=2.2]ccc@{}}
\toprule
SR & Comp. & {4DFlowNet} & FSR & PnP-ADMM & RAM\\
\midrule
$\times2$ & $u$ & 6.94 & 4.34 & 2.89 & \textbf{2.81}\\
& $v$ & 2.94 & 2.45 & 1.48 & \textbf{1.39}\\
& $w$ & 3.25 & 2.55 & 1.54 & \textbf{1.52}\\
\midrule
$\times4$ & $u$ & 14.42 & 6.99 & 6.53 & \textbf{5.58}\\
& $v$ & 6.56 & 3.26 & 2.87 & \textbf{2.54}\\
& $w$ & 7.20 & 3.63 & 3.25 & \textbf{2.88}\\
\bottomrule
\end{tabular*}
\end{table}

\begin{figure}[htbp]
\centerline{\includegraphics[width=\linewidth]{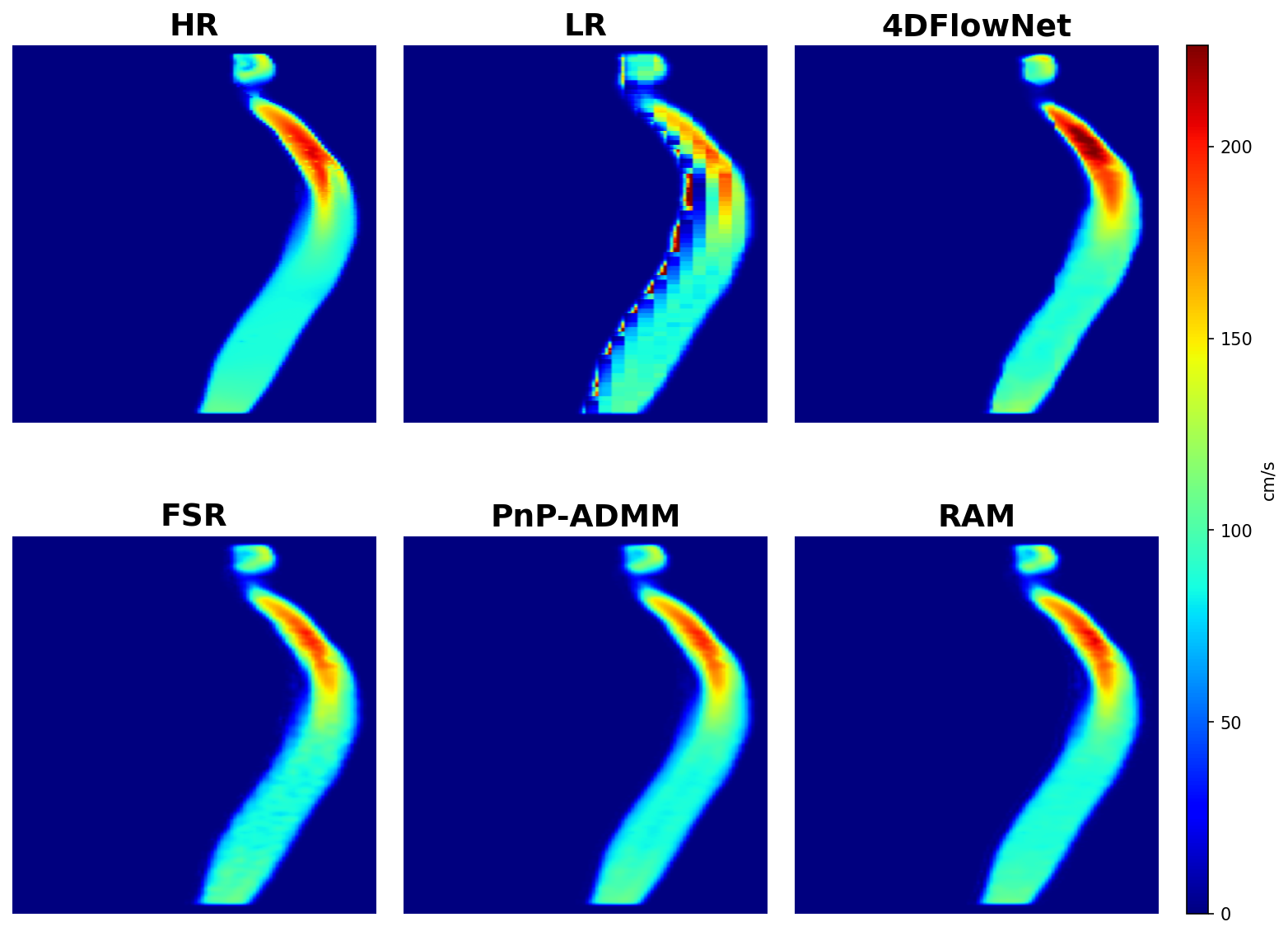}}
\caption{Sagittal slice of reconstructed velocity norm for the \texttt{aorta03} simulation dataset ($\times4$ super-resolution, time frame $t=12$). Values are masked to the vascular region.\label{fig:sr4-multislice-sagittal}}
\end{figure}

Figure \ref{fig:sr4-multislice-sagittal} shows the reconstructed velocity norm at $\delta=4$ on one sagittal slice, masked to display the vessel region only. In this frame, 4DFlowNet overestimates the velocity in the high-velocity region while FSR, PnP-ADMM and RAM do not. PnP-ADMM and RAM produce smoother maps than FSR inside the lumen.

Table \ref{tab:phantom-rmse} reports the masked RMSE per plane and vascular cross-section through the vascular mask. PnP-ADMM and RAM achieve the lowest RMSE in eight of the nine cases, ahead of FSR and 4DFlowNet.

\begin{table}[t]
\caption{Masked RMSE (cm/s) on the acquired aortic phantom ($\times2$ super-resolution) over 50 cardiac time frames, by plane and vascular cross-section (C1/C2). Bold marks the lowest error per row.}
\label{tab:phantom-rmse}
\centering\small
\setlength{\tabcolsep}{3pt}
\begin{tabular*}{\linewidth}{@{\extracolsep{\fill}}lccccc@{}}
\toprule
Plane & Cross-section & 4DFlowNet & FSR & PnP-ADMM & RAM\\
\midrule
AAO & C1 & 6.05 & 5.88 & 5.05 & \textbf{5.02}\\
& C2 & 3.77 & 4.44 & 2.98 & \textbf{2.93}\\
BCT & C1 & 4.87 & 4.21 & \textbf{4.03} & 4.14\\
& C2 & \textbf{2.39} & 4.23 & 2.81 & 2.98\\
BRANCH & C1 & 9.42 & 6.70 & \textbf{5.74} & 5.87\\
& C2 & 5.60 & 5.24 & 4.81 & \textbf{4.78}\\
INLET & C1 & 7.44 & 6.83 & 6.05 & \textbf{6.01}\\
& C2 & 3.97 & 4.31 & \textbf{3.12} & 3.26\\
LSA & C1 & 4.43 & 4.73 & 3.94 & \textbf{3.81}\\
\bottomrule
\end{tabular*}
\end{table}

\begin{figure}[htbp]
\centerline{\includegraphics[width=\linewidth]{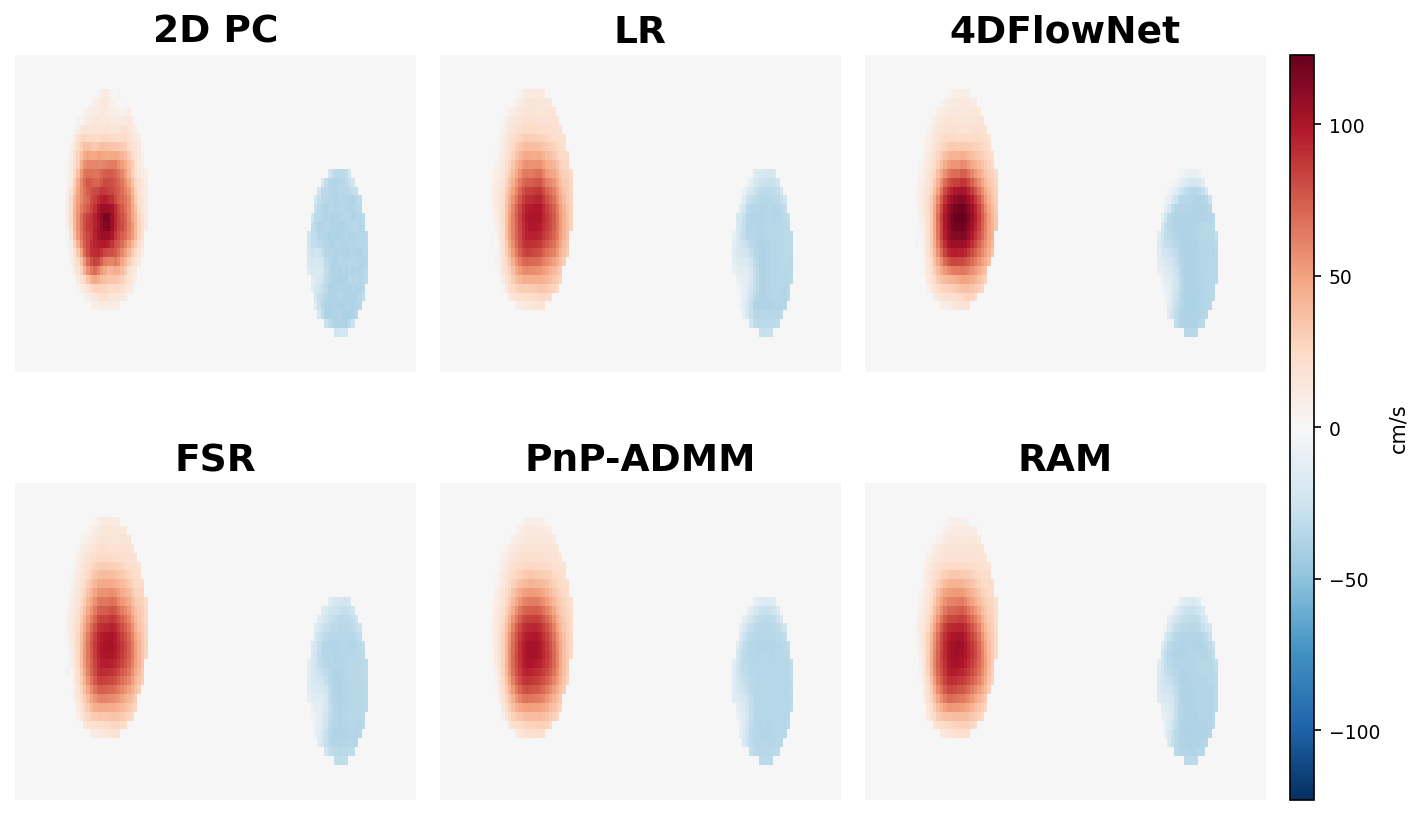}}
\caption{Phantom INLET plane perpendicular velocity ($\times2$ super-resolution, time frame $t=15$). Cross-section C1 is shown in red and C2 in blue. Values are masked to the vascular region.\label{fig:phantom}}
\end{figure}

Figure \ref{fig:phantom} shows, at a fixed anatomical plane, the 2D PC reference and the four reconstructions at one instant of the cardiac cycle. The plane intersects two vessels carrying flow in opposite directions. At this instant of peak flow, 4DFlowNet visibly overestimates velocities compared to the reference, while the other reconstructions remain closer to it.

These results support both the relevance of the proposed complex-domain formulation and the benefit of solving it by combining a model-based approach with pretrained networks, as shown by the results of PnP-ADMM and RAM. Nevertheless, some limitations remain. First, parameter selection for FSR and PnP-ADMM relies on the simulation ground truth and is reused unchanged on the phantom. Second, unlike 4DFlowNet, which is trained to output zero velocity outside the vessel, the inverse-problem methods do not suppress background noise. Third, validation is limited to one CFD geometry and one phantom with 2D reference measurements, so extended evaluation across anatomies, protocols and pathologies is required for stronger validation.

\section{Conclusion}\label{sec:conclusion}
We formulated spatial 4D Flow MRI super-resolution as a linear inverse problem on complex-valued signals using an operator inspired by MRI acquisition principles. This study demonstrates the potential of the proposed approach. Without task-specific training, FSR, PnP-ADMM, and RAM outperform 4DFlowNet on all tested simulation cases. PnP-ADMM and RAM also achieve a lower RMSE than 4DFlowNet on the phantom in eight of nine cases. Future work will address ground-truth-free parameter selection, background denoising, validation across additional anatomies and acquisitions, and the extension of all the methods to a fully 3D framework. Also, the competitiveness of PnP-ADMM motivates further work combining variational reconstruction with pretrained denoisers.

\section{COMPLIANCE WITH ETHICAL STANDARDS}
This research study was conducted on open-access simulation and phantom data for which no ethical approval was required.

\section{CONFLICTS OF INTEREST}
No funding was received for conducting this study and the authors have no relevant financial or non-financial interests to disclose.

\bibliographystyle{IEEEbib}
\bibliography{strings,refs}
\end{document}